# Sequential Recommender Systems: Challenges, Progress and Prospects *

Shoujin Wang[1], Liang Hu[2], Yan Wang[1], Longbing Cao[2], Quan Z. Sheng[1], Mehmet Orgun[1]

[1]Department of Computing, Macquarie University

[2]Advanced Analytics Institute, University of Technology Sydney

{shoujin.wang, yan.wang}@mq.edu.au, rainmilk@gmail.com,

## Abstract

The emerging topic of sequential recommender systems (SRSs) has attracted increasing attention in recent years. Different from the conventional recommender systems (RSs) including collaborative filtering and content-based filtering, SRSs try to understand and model the sequential user behaviors, the interactions between users and items, and the evolution of users' preferences and item popularity over time. SRSs involve the above aspects for more precise characterization of user contexts, intent and goals, and item consumption trend, leading to more accurate, customized and dynamic recommendations. In this paper, we provide a systematic review on SRSs. We first present the characteristics of SRSs, and then summarize and categorize the key challenges in this research area, followed by the corresponding research progress consisting of the most recent and representative developments on this topic. Finally, we discuss the important research directions in this vibrant area.

## 1 Introduction

Sequential recommender systems (SRSs) suggest items which may be of interest to a user by mainly modelling the sequential dependencies over the user-item interactions (e.g., view or purchase items on an online shopping platform) in a sequence [27]. The traditional recommender systems (RSs), including the content-based and collaborative filtering RSs, model the user-item interactions in a static way and can only capture the users' general preferences. In contrast, SRSs treat the user-item interactions as a dynamic sequence and take the sequential dependencies into account to capture the current and recent preference of a user for more accurate recommendation [1]. In order to enhance the understanding of SRSs, next we present the motivation and formalization of SRSs.

---

A comprehensive survey on session-based recommender systems can be found: https://arxiv.org/abs/1902.04864.

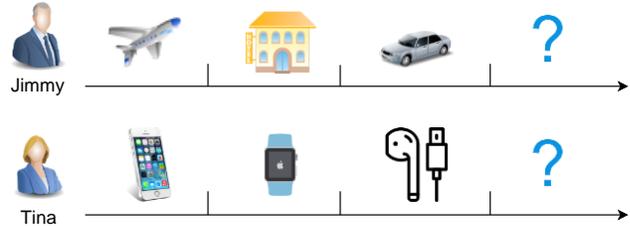

Figure 1: Two examples of SRSs: (1) After Jimmy has booked a flight, a hotel and rented a car, what will be his next action? (2) After Tina has bought an iPhone, an iWatch and a pair of AirPods, what would she buy next?

Motivation: Why Sequential Recommender Systems?

*The user-item interactions are essentially sequentially dependent.* In the real world, users' shopping behaviours usually happen successively in a sequence, rather than in an isolated manner. Taking the shopping events of Jimmy depicted in Figure 1 as an example, before Jimmy started holiday, he booked a flight and a hotel and rented a car successively, and his next action may be visiting a tourist attraction via selfdriving. In such a case, the hotel may be close to the destination airport of the flight and the location for picking up the rented car may be not far away from the hotel. In this scenario, each of Jimmy's next actions depends on the prior ones and thus all the four consumption actions are sequentially dependent. Likewise, we can see the sequential dependencies in Tina's case. Such kind of sequential dependencies commonly exist in transaction data but cannot be well captured by the conventional content-based RSs or collaborative filtering RSs [12], which essentially motivates the development of SRSs.

*Both the users' preference and items' popularity are dynamic rather than static over time.* In fact, a user's preference and taste may change over time. For instance, many young people who used to be iPhone fans now have switched to become fans of the phones manufactured by Huawei or Samsung and the popularity of iPhone has been dropping in recent years. Such dynamics are of great significance for precisely profiling a user or an item for more accurate recommendations and they can only be captured by SRSs.

*User-item interactions usually happen under a certain sequential context.* Different contexts usually lead to different users' interactions with items, which is, however, often ignored by traditional RSs like collaborative filtering. In contrast, an SRS takes the prior sequential interactions as a context to predict which items would be interacted in the near future. As a result, it is much easier to diversify the recommendation results by avoiding repeatedly recommending those items identical or similar to those already chosen.

Formalization: What are Sequential Recommender Systems?

Generally, an SRS takes a sequence of user-item interactions as the input and tries to predict the subsequent user-item interactions that may happen in the near future through modelling the complex sequential dependencies embedded in the sequence of user-item interactions. More specifically, given a sequence of user-item interactions, a recommendation list consisting of top ranked candidate items are generated by maximizing a utility function value (e.g., the likelihood):

$$R = arg\ max\ f(S) \qquad (1)$$

where *f* is a utility function to output a ranking score for the candidate items, and it could be of various forms, like a conditional probability [19], or an interaction score [11]. $S = \{i_1, i_2, ..., i_{|S|}\}$ is a sequence of user-item interactions where each interaction $i_j = <u, a, v>$ is a triple consisting of a user *u*, the user's action *a*, and the corresponding item *v*. In some cases, users and items are associated with some meta data (e.g., the demographics or the features), while the actions may have different types (e.g., click, add to the cart, purchase) and happen under various contexts (e.g., the time, location, weather). The output *R* is a list of items ordered by the ranking score.

Different from the general sequence modelling in which the sequence structure is much simpler since a sequence is often composed of atomic elements (e.g., real values, genes), the learning task in SRSs is much more challenging because of the more complex sequence structure (e.g., each element is a triple). This motivates us to systematically analyze the challenges in SRSs and summarize the corresponding progress. Contributions. The main contributions of this work are summarized below:

- We systematically analyze a number of key challenges caused by different data characteristics in SRSs and categorize them from a data driven perspective, which provides a new view to deeply understand the characteristics of SRSs.
- We summarize the current research progress in SRSs by systematically categorizing the state-of-the-art works from a technical perspective.
- We share and discuss some prospects of SRSs for the reference of the community.

## 2 Data Characteristics and Challenges

Due to the diversity and complexity of the customers' shopping behaviours, item characteristics and the specific shopping contexts in the real world, the generated user-item interaction data often has different characteristics. Different data characteristics essentially bring different challenges for SRSs, which require different solutions, as presented in Table 1. In the following five subsections, we specifically discuss five key challenges respectively in SRSs caused by different data characteristics. In each subsection, we first introduce the particular data characteristics and then illustrate the corresponding challenges.

### 2.1 Handling Long User-Item Interaction Sequences

A long user-item interaction sequence consists of a relatively large number of user-item interactions. As a result, it has a much higher chance to have more complex and comprehensive dependencies over the multiple interactions inside it, which makes the sequential recommendations much more challenging. Specifically, two most critical challenges in long user-item interaction sequences are *learning higher-order sequential dependencies* and *learning long-term sequential dependencies*, which will be presented respectively below. Learning higher-order sequential dependencies. Hig herorder sequential dependencies commonly exist in the useritem interaction sequences, especially in long ones. Compared to the lower-order sequential dependencies, which are relatively simple and can be easily modeled by Markov chain models [3] or factorization machines [14; 10], higher-order sequential dependencies are much more complex and harder to be captured because of their complicated multi-level cascading dependencies crossing multiple user-item interactions. So far, there have been mainly two basic approaches reported that can address this challenge in SRSs to some extent: *higher-order Markov-chain models* [6] and *recurrent neural networks (RNN)* [7], as shown in Table 1. However, each approach has its own limitations, for example, the historical states that can be involved in a higher-order Markov-chain model are quite limited as the number of the model parameters to be estimated grows exponentially with the order, while the overly strong order assumption employed in RNN limits the application of RNN in sequences with a flexible order. The technical progress achieved in both approaches will be presented in Sections 3.1 and 3.3 respectively in more details. Learning long-term sequential dependencies. Long-term sequential dependencies refer to the dependencies between interactions that are far from each other in a sequence. For instance, given a shopping sequence $S_1 = \{a\ rose,\ eggs,\ bread,\ a\ bottle\ of\ milk,\ a\ vase\}$, which consists of a basket of items that are purchased successively by a user Janet. Obviously, the vase and the rose are highly dependent even though they are far from each other. Such cases are not uncommon in the real world as users' behaviours are usually highly uncertain and thus they may put any items into the cart. To address such a critical issue, Long Short Term Memory (LSTM)-based [21] and Gated Recurrent Unit (GRU)-based [7] RNN have been applied in SRSs to capture the long-term dependencies among the user-item interactions in a sequence. However, it is easy for RNN models to generate false dependencies by overly assuming any adjacent items in a sequence are highly dependent. In the above example of Janet's shopping sequence, an RNN usually models $S_1$ by assuming the milk and vase are dependent due to the close distance between them, but actually they are not. Some other efforts have been made to solve this issue by utilizing the advantage of mixture models to combine multiple sub-models with different temporal ranges to capture both short- and long-term dependencies in a unified model [15]. Overall, the works that are able to tackle this challenge are quite limited and more investigations are required to bridge this gap. The technical progress achieved in RNN and mixture models will be presented in Section 3.3.

## 2.2 Handling User-Item Interaction Sequences with a Flexible Order

In the real world, some user-item interaction sequences are strictly ordered while others may not be, namely, not all adjacent interactions are sequentially dependent in a sequence [4]. For instance, in a shopping sequence $S_2$ = {milk, butter, flour}, it does not matter whether to buy milk or butter first, but the purchase of both items leads to a higher probability of buying flour next; namely, there is no strict order between milk and butter, but flour sequentially depends on the union of them. Therefore, for a sequence with a flexible order, it is much better to capture the *collective sequential dependencies*, rather than the point-wise ones as the former is fuzzy and does not assume a strict order over user-item interactions. As a result, *how to capture collective sequential dependencies under an assumption of flexible order* becomes the key challenge in handling sequences with a flexible order in SRSs.

Although common and important, reported studies in SRSs have not paid much attention to this issue yet. Existing SRSs built on Markov-chains, factorization machines or RNN can only handle the point-wise dependencies but are not good at modelling and capturing collective dependencies. Only quite few works like [17; 26] have attempted to address such a challenge by employing the strength of convolutional neural networks (CNN) to model the local and global dependencies between different areas in an "image", i.e., the embedding matrix of a sequence of interactions. The technical progress achieved in CNN-based SRSs will be presented in Section 3.3.

## 2.3 Handling User-Item Interaction Sequences with Noise

Due to the uncertainty of user shopping behaviours, most of the user-item interaction sequences are not clean, meaning that they may contain some noisy and irrelevant interactions that generate interference for the next interaction prediction. In practice, in a user-item interaction sequence, some historical interactions are strongly relevant to the next interaction, while others may be weakly relevant or even irrelevant. For example, in another shopping sequence $S_3$ = {bacon, a rose, eggs, bread}, the item "rose" may be a noisy item as it is quite different from others and has no correlation to them. The next item may be a bottle of milk with a high probability and it only sequentially depends on bacon, eggs and bread while has nothing to to with the rose. Therefore, another key challenge in SRSs is *to learn sequential dependencies attentively and discriminatively* over user-item interaction sequences with noise.

Quite a few works have attempted to solve such a typical issue by employing the attention models [19] or memory networks [1] to selectively retain and utilize information from those interactions that are truly relevant to the next interaction prediction. The technical progress achieved in these solutions will be presented in Section 3.3.

## 2.4 Handling User-Item Interaction Sequences with Heterogeneous Relations

Heterogeneous relations refer to different types of relations which deliver different kinds of information and should be modelled differently in SRSs. For instance, in a user-item interaction sequence, except for the widespread occurrencebased sequential dependencies over user-item interactions, there are also similarity-based relations between the interacted items in terms of their features. Furthermore, even though both are sequential dependencies, long-term sequential dependencies are quite different from short-term ones and they cannot be modelled in the same way. Therefore, another key challenge in SRSs is *how to effectively capture these heterogeneous relations embedded in the user-item interaction sequences respectively and to make them work collaboratively for the sequential recommendations* when handling user-item interaction sequences associated with heterogeneous relations.

There are quite limited works reported in the literature to solve this challenge in SRSs. Mixture models [12; 15; 20] are the only solution to address such challenge so far. A mixture model integrates different types of relations modelled by different sub-models to collaboratively generate sequential recommendations. The specific technical progress will be presented in Section 3.3.

## 2.5 Handling User-Item Interaction Sequences with Hierarchical Structures

Generally, there are mainly two kinds of hierarchical structures that may be associated with a user-item interaction sequence: (1) *the hierarchical structure between meta data and user-item interactions*. To be specific, the users' demographics can determine the users' preferences in some degree and can further affect their interactions with the items. Similarly, the features of items often have some effects on whether they will be liked and interacted by users [9]; and (2) *the hierarchical structure between subsequences and user-item interactions*. More specifically, in some SRSs, one user-item interaction sequence includes multiple sub-sequences (also called sessions). In such a case, in addition to the prior interactions within the current sub-sequence, the historical subsequences may also influence the next user-item interaction to be predicted in the current sub-sequence [25]. Therefore, one more key challenge in SRSs is *how to incorporate the hierarchical dependencies embedded in these two kinds of hierarchical structures into sequential dependency learning to generate more accurate sequential recommendations.*

Table 1: A summary of challenges driven by data characteristics in SRSs

| Data characteristics | Challenges | Existing solutions |
|---|---|---|
| Long user-item interaction sequences | Learning higher-order sequential dependencies | Higher-order Markov chain [He and McAuley, 2016], RNN [Hidasi et al., 2016a] |
| | Learning long-term sequential dependencies | LSTM- [Wu et al., 2017] and GRU-based [Hidasi et al., 2016a] RNN, mixture models [Tang et al., 2019] |
| User-item interaction sequences with a flexible order | Learning collective sequential dependencies under the assumption of flexible order | CNN [Tang and Wang, 2018; Yuan et al., 2019] |
| User-item interaction sequences with noise | Learning sequential dependencies attentively and discriminatively | Attention models [Wang et al., 2018], memory networks [Chen et al., 2018] |
| User-item interaction sequences with heterogeneous relations | Learning heterogeneous relations discriminatively and integrating them effectively | Mixture models [Kang et al., 2018; Tang et al., 2019] |
| User-item interaction sequences with hierarchical structures | Learning hierarchical dependencies | Feature-enriched RNN [Hidasi et al., 2016b], hierarchical embedding [Wang et al., 2015], hierarchical RNN [Quadrana et al., 2017], hierarchical attention [Ying et al., 2018] |

Although quite a few works have attempted to address this challenge from certain aspects, some other aspects have been less studied. On the one hand, to take the influences of items' features on the user-item interactions into account, a series of feature-enriched neural models including [9] have been proposed for SRSs. In comparison, the influences of users' demographics have been rarely considered in existing SRSs and more efforts should be devoted into this direction. On the other hand, some hierarchical models including hierarchical embedding models [18], hierarchical RNN [13] and hierarchical attention networks [25] have been devised to incorporate the historical sub-sequences into sequential dependency learning to build more powerful SRSs. Particularly, the technical progress achieved to address this challenge will be presented in Sections 3.2 and 3.3.

## 3 Research Progress

To provide an overview of the technical progress in SRSs and to give more technical details of the solutions to the aforementioned challenges, we summarize and briefly discuss the research progress in SRSs from a technical perspective in this section. Particularly, we first present a categorization of all the approaches for SRSs from the technical perspective and then briefly highlight the recent progress in each category.

The categorization of SRS approaches is presented in Figure 2. We observe that the various approaches for SRSs are first categorized into 11 atomic classes ( e.g., the sequential pattern mining, factorization machine, and recurrent neural networks) from the technical perspective. All these atomic classes are then further categorized into three taxonomies, including traditional sequence models, latent representation models, and deep neural network models. Generally speaking, these three taxonomies change from simple to complex and are reported successively. Next we summarize the research progress in each of the three taxonomies.

### 3.1 Traditional Sequence Models for SRSs

Traditional sequence models including sequential pattern mining and Markov chain models are intuitive solutions to SRSs by taking advantage of their natural strength in modelling sequential dependencies among the user-item interactions in a sequence.

Sequential pattern mining. Sequential pattern-based RSs first mine frequent patterns on sequence data and then utilize the mined patterns to guide the subsequent recommendations. Although simple and straightforward, sequential pattern mining usually generates a large number of redundant patterns, which increases unnecessary cost w.r.t. time and space. Another obvious shortcoming is that it often loses those infrequent patterns and items due to the frequency constraint, which limits the recommendation results to those popular items. Therefore, quite few works have been reported in this class, except a representative one [24; 16].

Markov chain models. Markov chain-based RSs adopt Markov chain models to model the transitions over user-item interactions in a sequence, for the prediction of the next interaction. According to the specific technique used, Markov chain-based RSs are divided into *basic Markov Chain-based approaches* and *latent Markov embedding-based approaches*. The former one directly calculates the transition probability based on the explicit observations [3], while the latter first embeds the Markov chains into an Euclidean space and then calculates the transition probabilities between interactions based on their Euclidean distance [2]. The shortcomings of Markov chain-based RSs are obvious, namely, on the one hand, they can only capture the short-term dependencies while ignoring long-term ones due to the Markov property which assumes that the current interaction depends on one or several most recent interactions only; on the other hand, they can only capture the point-wise dependencies while ignoring the collective dependencies over user-item interactions. Consequently, they are less and less employed in SRSs in recent years.

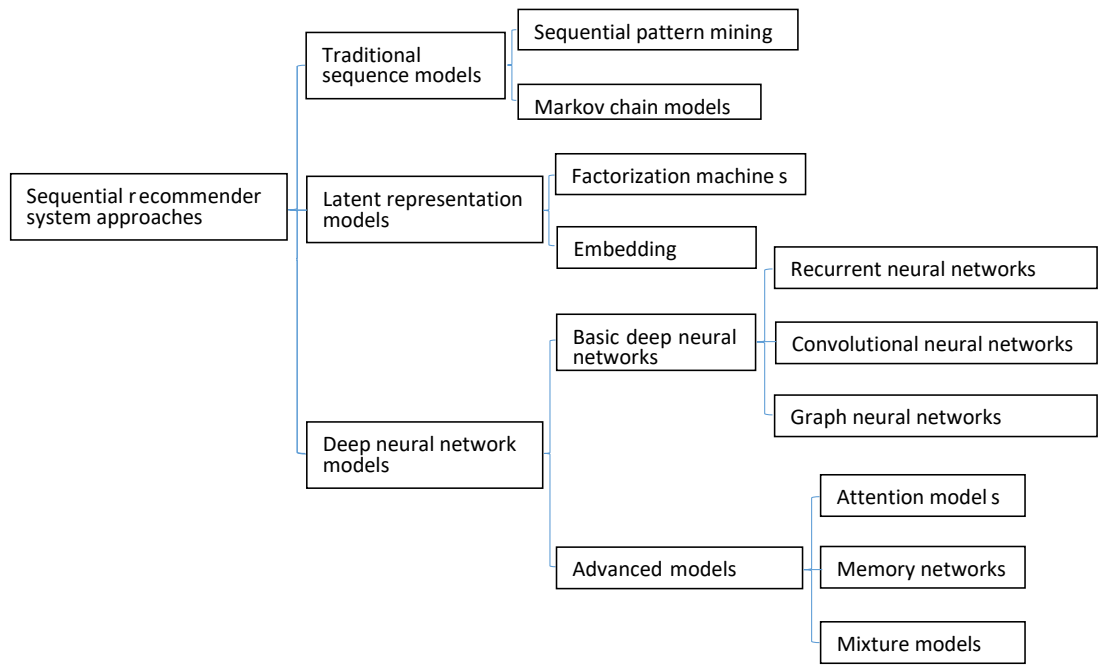

Figure 2: A categorization of SRS approaches from the technical perspective

## 3.2 Latent Representation Models for SRSs

Latent representation models first learn a latent representation of each user or item, and then predict the subsequent useritem interactions by utilizing the learned representations. As a result, more implicit and complex dependencies are captured in a latent space, which greatly benefits the recommendations. Next, we introduce two representative models falling into this taxonomy.

Factorization machines. Factorization machine-based SRSs usually utilize the matrix factorization or tensor factorization to factorize the observed user-item interactions into latent factors of users and items for recommendations [14; 10]. Different from collaborative filtering (CF), the matrix or tensor to be factorized is composed of interactions rather than the ratings in CF. Such a model is easily affected by the sparsity of the observed data and thus cannot achieve ideal recommendations.

Embedding. Embedding-based SRSs learn a latent representations for each user and item for the subsequent recommendations by encoding all the user-item interactions in a sequence into a latent space. Specifically, some works take the learned latent representations as the input of a network to further calculate an interaction score between users and items, or successive users' actions [18; 19], while other works directly utilize them to calculate a metric like the Euclidean distance as the interaction score [5]. This model has shown great potential in recent years due to its simplicity, efficiency and efficacy.

## 3.3 Deep Neural Network Models for SRSs

Deep neural networks [23] have natural strength to model and capture the comprehensive relations over different entities (e.g., users, items, interactions) in a sequence, and thus they nearly dominate SRSs in the past few years. The latest progress achieved in SRSs also belongs to this taxonomy. Generally, this taxonomy can be divided into two sub classes: SRSs built on basic deep neural networks and SRSs built on deep neural networks with some advanced models incorporated.

Basic Deep Neural Networks
The most commonly used deep neural networks for SRSs are recurrent neural networks (RNN) due to their natural strength in sequence modelling, but they also have defects.

Recently, convolutional neural networks (CNN) and graph neural networks (GNN) have also been applied in SRSs to make up the defects of RNN. Next, we introduce the SRSs built on top of these three deep neural networks respectively.

RNN-based SRSs. Given a sequence of historical user-item interactions, an RNN-based SRS tries to predict the next possible interaction by modelling the sequential dependencies over the given interactions. Except for the basic RNN, longshort-term-memory (LSTM)- [21] and gated recurrent unit (GRU)-based [7] RNN have also been developed to capture the long-term dependencies in a sequence. Recent years have witnessed the prosperity of RNN-based SRSs and they dominate the research on the deep learning-based SRSs or even the whole SRSs. Besides the basic RNN structure, some variants are proposed to capture more complex dependencies in a sequence, like hierarchical RNN [13]. However, RNN is not flawless for SRSs, with the shortcomings in two aspects: (1) it is easy to generate fake dependencies due to the overly strong assumption that any adjacent interactions in a sequence must be dependent, which may not be the cases in the real world because there are usually irrelevant or noisy interactions inside a sequence; and (2) it is likely to capture the point-wise dependencies only while ignoring the collective dependencies (e.g., several interactions collaboratively affect the next one).

CNN-based SRSs. Different from RNN, given a sequence of user-item interactions, a CNN first puts all the embeddings of these interactions into a matrix, and then treats such a matrix as an "image" in the time and latent spaces. Finally, a CNN learns sequential patterns as local features of the image using convolutional filters for the subsequent recommendations. Since a CNN does not have strong order assumptions over the interactions in a sequence, and they learn patterns between the areas in an "image" rather than over interactions, therefore, CNN-based SRSs can make up the aforementioned drawbacks of RNN-based SRSs to some degree. However, CNN-based SRSs cannot effectively capture long-term dependencies due to the limited sizes of the filters used in CNN, which limits their applications. The typical works include [17; 26].

GNN-based SRSs. Recently, with the fast development of GNN, GNN-based SRSs have been devised to leverage GNN to model and capture the complex transitions over user-item interactions in a sequence. Typically a directed graph is first built on the sequence data by taking each interaction as a node in the graph while each sequence is mapped to a path. Then, the embeddings of users or items are learned on the graph to embed more complex relations over the whole graph [22]. Such an approach makes full use of the advantage of GNN to capture the complex relations in structured relation datasets. GNN-based SRSs have shown a great potential to provide explainable recommendations by revealing the complex relations between the recommended items and the corresponding sequential context. Such kind of SRSs are still in their early stages.

Advanced Models

To address the limitations of SRSs built on basic neural network structures, some advanced models are usually combined together with a certain kind of basic deep neural networks (e.g., RNN, CNN) to build more powerful SRSs which are able to address particular challenges. Next, we introduce three advanced models that are commonly used in SRSs.

Attention models. Attention models are commonly employed in SRSs to emphasize those really relevant and important interactions in a sequence while downplaying those ones irrelevant to the next interaction. They are widely incorporated into shallow networks [19] and RNN [25] to handle interaction sequences with noise.

Memory networks. Memory networks are introduced into SRSs to capture the dependencies between any historical user-item interaction and the next one directly by incorporating an external memory matrix. Such matrix enables it possible to store and update the historical interactions in a sequence more explicitly and dynamically to improve the expressiveness of the model and reduce the interference of those irrelevant interactions [1]. Furthermore, some works incorporate a key-value memory network to store and update the corresponding knowledge base information of the interacted items in a sequence to learn the attribute level preference for the enhancement of recommendations [11]. Generally, memory networks have shown their potential in SRSs, but are not sufficiently studied yet.

Mixture models. A mixture model-based SRS combines different models that excel at capturing different kinds of dependencies to enhance the capability of the whole model in capturing various dependencies for better recommendations. A typical example is [15], which combines different kinds of encoders that are suitable for short- and long-term dependencies respectively to learn a more precise sequence representation for the subsequent recommendations and has demonstrated to be quite effective. However, such models are in their early stages.

## 4 Open Research Directions

Recent years, particularly the recent three years, have witnessed the fast development of sequential recommender systems, along with the prosperity of deep learning, especially that of the recurrent neural networks. While categorizing and summarizing the research practices in this filed, we have identified further open research directions discussed below.

Context-aware sequential recommender systems. The current context in which a user or an item is could greatly influence the user's choice on the item, which should be considered when conducting recommendations. This is even more necessary in SRSs as the context may change over time.

However, most existing SRSs ignore such significant aspect. Therefore, context-aware SRSs would be an important direction for future work.

Social-aware sequential recommender systems. Users live in a society and are connected with various people both online and offline. Others' behaviors or opinions often affect the users' choices greatly. Therefore, the social influence needs to be taken into account in SRSs, which is usually ignored in the existing works.

Interactive sequential recommender systems. Most of shopping behaviours in the real-world are continuous rather than isolated events. In other words, there are actually sequential interactions between a user and the shopping platform (e.g., Amazon). However, the existing SRSs often neglect such interactions and only generate recommendations for one action at a single time step. How to incorporate the user-seller interactions and thus generate multi-time step recommendations is a promising research direction.

Cross-domain sequential recommender systems. In the real world, items purchased by a user during a certain time period are often from multi-domains rather than one domain. Essentially, there are some sequential dependencies between items from different domains, such as the purchase of a car insurance after the purchase of a car. Such cross-domain sequential dependencies are ignored in most SRSs. Therefore, cross-domain SRS is another promising research direction to generate more accurate recommendations by leveraging information from other domains and more diverse recommendations from different domains.

## 5 Conclusions

Recommender systems (RS) is one of the most direct and practical applications of artificial intelligence in our daily lives. Sequential recommender systems (SRSs) have been at the core of the RS field in the past three to five years as they provide more intelligent and favorable recommendations to satisfy our daily requirements. It is our hope that this summary provides an overview of the challenges and the recent progress as well as some future directions in SRSs to the RS research community.

## Acknowledgements

This work was partially supported by Australian Research Council Discovery Project DP180102378.